\documentclass[prb,aps,unsortedaddress,superscriptaddress,endfloats]{revtex4}
\usepackage{amsmath}
\usepackage{amssymb}
\usepackage{graphicx}
\begin{document}
\title{X ray emission line profile modeling of hot stars}
\author{Roban H. Kramer}
\affiliation{Swarthmore College, Swarthmore, PA, 19081}
\affiliation{Prism Computational Sciences, Madison, WI 53711}
\author{Stephanie K. Tonnesen}
\affiliation{Swarthmore College, Swarthmore, PA, 19081}
\author{David H. Cohen}
\affiliation{Swarthmore College, Swarthmore, PA, 19081}
\affiliation{Prism Computational Sciences, Madison, WI 53711}
\author{Stanley P. Owocki}
\author{Asif ud-Doula}
\affiliation{Bartol Research Institute, University of Delaware, Newark, DE 19716}
\author{Joseph J. MacFarlane}
\affiliation{Prism Computational Sciences, Madison, WI 53711}

\begin{abstract}
The launch of high-spectral-resolution x-ray telescopes
(\textit{Chandra}, \textit{XMM}) has provided a host of new spectral
line diagnostics for the astrophysics community.  In this paper we
discuss Doppler-broadened emission line profiles from highly
supersonic outflows of massive stars.  These outflows, or winds, are
driven by radiation pressure and carry a tremendous amount of kinetic
energy, which can be converted to x rays by shock-heating even a small
fraction of the wind plasma.  The unshocked, cold wind is a source of
continuum opacity to the x rays generated in the shock-heated portion
of the wind.  Thus the emergent line profiles are affected by
transport through a two-component, moving, optically thick medium.
While complicated, the interactions among these physical effects can
provide quantitative information about the spatial distribution and
velocity of the x-ray-emitting and absorbing plasma in stellar winds.
We present quantitative models of both a spherically-symmetric wind
and a wind with hot plasma confined in an equatorial disk by a dipole
magnetic field.
\end{abstract}
\maketitle

\section{Introduction}
Ultraviolet spectra of O and B stars (with luminosities up to $L=10^6
\;\mathrm{L_{\odot}}$ and surface temperatures $T \gtrsim 3
\;\mathrm{T_{\odot}}$) show the signatures of rapidly expanding winds,
with velocities on the order of a few $1000\; \mathrm{km}\;
\mathrm{s}^{-1}$, densities of order $10^{10}\; \mathrm{cm}^{-3}$, and
mass-loss rates up to $10^{-5}\;\mathrm{M_{\odot}
\;\mathrm{yr}}^{-1}$. These stars are detected to have soft-x-ray
luminosities of $~10^{-7}$ times their total (bolometric)
luminosities.\cite{1989ApJ...341..427C} In cooler stars, like the Sun,
x rays are produced in coronae, which are high-temperature regions
near the surface of the star, and their x-ray line widths are
primarily indicative of thermal broadening. The line widths observed
in x-ray spectra of hot stars, however, provide evidence that the
emission originates in extended, high-velocity
winds. \cite{2001A&A...365L.312K}

The \textit{Chandra X-ray Observatory} has the spectral resolution to
resolve Doppler-broadened lines from hot stars, providing quantitative
information about the geometry and kinematics of the wind and putting
important constraints on physical models that may explain the observed
x rays. \textit{Chandra} uses a set of concentric parabolic and
hyperbolic mirrors with an effective area of about $10\;
\mathrm{cm}^2$ to focus x rays on its instrument package. The Medium
Energy Grating (MEG) provides a resolution of $\Delta \lambda =
0.023\;\mathrm{\mbox{\AA}}$ from $0.4$ to $5.0\;\mathrm{keV}$ ($31$ to
$2.5\;\mathrm{\mbox{\AA}}$).

\section{Spherically-Symmetric Wind Models}

An optically-thin, spherically-expanding, x-ray-emitting plasma
produces symmetric line profiles centered on the rest wavelengths of
the lines with a shape determined entirely by the velocity and
emissivity structure of the plasma. Two effects can change the symmetry
of the line. The star itself occults part of the plasma, while cold
components of the wind absorb x rays, meaning photons traveling along
longer lines of sight or through more dense wind regions will be less
likely to reach the observer. In a radially expanding wind, the
observer sees blue-shifted emission from the approaching material on
the near side of the wind and red-shifted emission from the receding
material on the far side. Thus an optically thick wind has the effect
of removing photons from the red side of the profile, shifting the
line center blueward and altering the shape of the line.

Owocki and Cohen \cite{2001ApJ...559.1108O} have numerically
integrated a spherically-symmetric wind model to determine the shape
of line profiles produced in a wind characterized by four
parameters. The wind velocity at a distance $r$ from the center of the
star is assumed to be of the form $v(r) = v_\infty (1-R_*/r)^\beta$ (a
``$\beta$ velocity law''), where $R_*$ is the stellar radius,
$v_\infty$ is the terminal velocity of the wind (determined from
ultraviolet spectra), and $\beta$ is a free parameter of order
unity. Given a constant mass-loss rate, the density at any point in a
smoothly-expanding, spherically-symmetric wind can be determined from
the velocity law. The emissivity of the wind is taken to be zero below
some radius $R_0$, and to fall off like $n^2 r^{-q}$ above that, where
$n$ is the density of the wind and $q$ is a free parameter. The
optical depth of the wind is characterized by the parameter $\tau_*$,
which is defined so that, in a constant velocity wind ($\beta = 0$,
$v=v_\infty$), $\tau_*$ is the radial optical depth at the surface of
the star and the radius of optical depth unity is $r|_{\tau=1}= \tau_*
R_*$. The parameters $\beta$, $q$, $R_0$, and $\tau_*$ offer enough
flexibility to characterize a variety of physical models, including
coronal models (by setting $q$ and $\beta$ to high values and
$R_0=R_*$).

All lines produced in an extended wind of this form will be skewed to
the blue side of line center by the removal of photons from the red
edge of the profile. \textit{Chandra} observations of the star
$\mathrm{\zeta}$ Puppis \cite{2001ApJ...554L..55C} reveal broad lines
with a significant blueward skew, suggesting an extended, spherical
wind model may be able to explain the shape of the profiles.

\begin{figure}
\includegraphics[width=8.5cm,height=10.9cm]{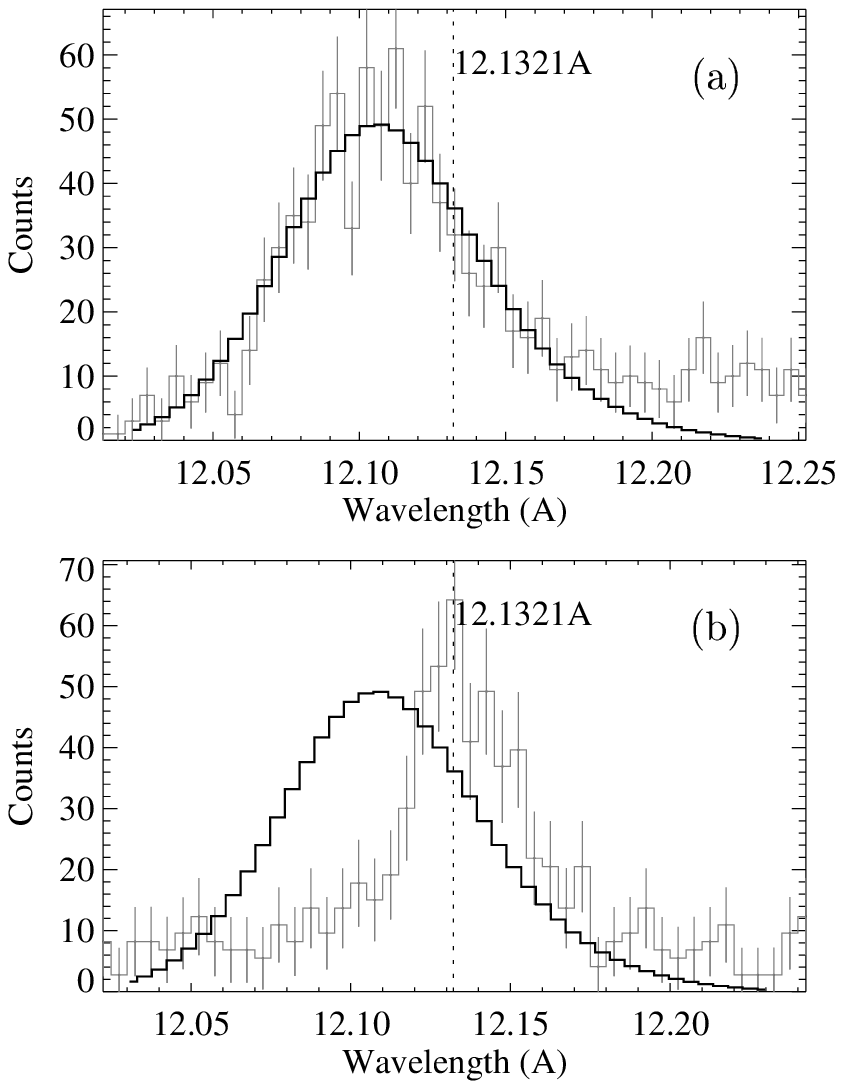}
\caption{\label{robanpic} \textit{Chandra} MEG ($+1$ and $-1$ order
summed) spectra (with error bars) of the Ne \textsc{X} Lyman $\alpha$
line in (a) $\mathrm{\zeta}$ Puppis, and (b)
$\mathrm{\mathrm{\theta}}^1$ Orionis C, with the best-fit
spherically-symmetric wind model for $\mathrm{\zeta}$ Puppis (in
black). The parameters for the best-fit model are $\beta=1$, $q = 1.2$,
$R_0 = 1.1 R_*$, and $\tau_* = 2.5$. The area under the model curve is
normalized to the area under the data curve.}
\end{figure}

Preliminary work on fitting the Owocki and Cohen model to lines in
$\mathrm{\zeta}$ Puppis suggest that good fits may be obtained with
physically reasonable parameters. Figure \ref{robanpic} (a) shows the
\textit{Chandra} MEG (sum of $+1$, and $-1$ order) spectrum of the
$12.13\;\mathrm{\mbox{\AA}}$ Ne \textsc{X} Lyman $\alpha$ line of $\mathrm{\zeta}$
Puppis along with the best-fit line profile.

Other hot stars, however, show much more symmetric profiles. Figure
\ref{robanpic} (b) shows the same Ne \textsc{X} line of
$\mathrm{\theta}^1$ Orionis C along with the model that fit the line
in $\mathrm{\zeta}$ Puppis. No spherically-symmetric wind model that
includes absorption can fit the relatively narrow and symmetric line
from this star.

\section{Modeling non-spherically-symmetric winds}

A promising alternative to spherically-symmetric models is the
Magnetically Confined Wind Shock (MCWS) model in which a strong (kG),
large-scale dipole magnetic field originates from the star's surface
and directs the stellar outflow.  This field directs the wind toward
the magnetic equator, where flows from opposite hemispheres collide,
causing strong shocks and heating the wind plasma to $T \approx 10^7$ K.  This
theory, originally proposed by Babel and Montmerle
\cite{1997A&A...323..121B}, lately has gained currency due to the
direct detection of an 1100 G field on $\mathrm{\mathrm{\theta}}^1$
Orionis C \cite{2002MNRAS.333...55D} and the detailed
magnetohydrodynamic modeling by ud-Doula and Owocki.
\footnote{A. ud-Doula and S. P. Owocki. Astrophys. J. (in press,
accepted 2 May 2002)}

\begin{figure}
\includegraphics[width=8.5cm,height=16.5cm]{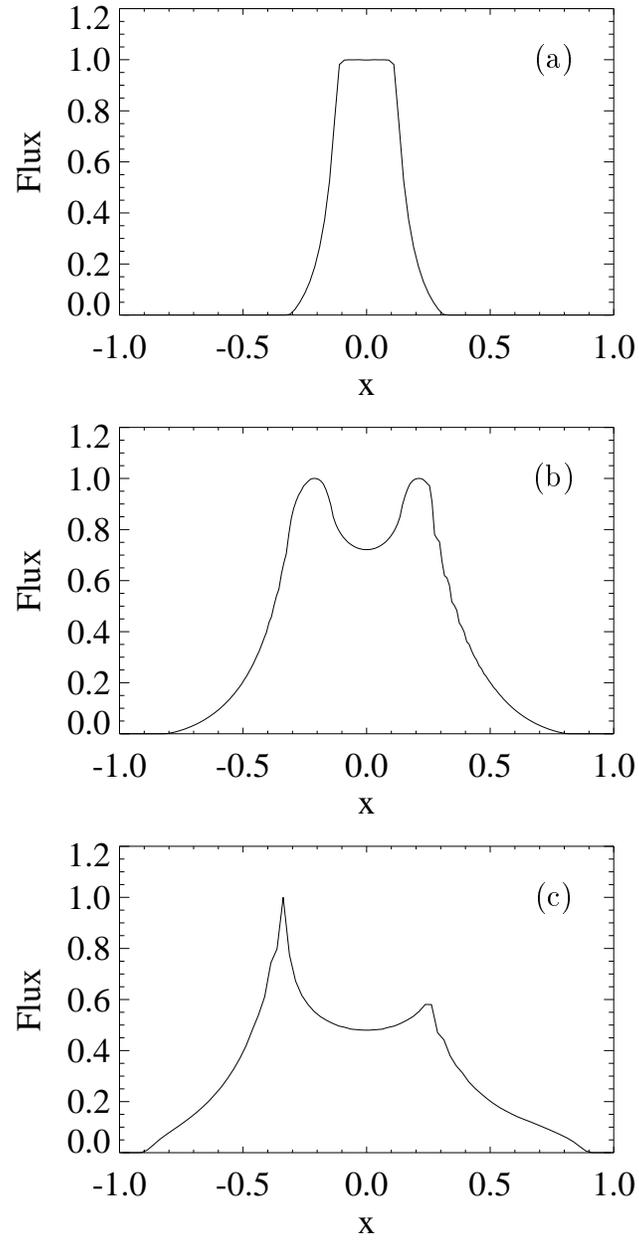}
\caption{\label{stephpic} Line profiles of an equatorial disk for an
observer at three different viewing angles (a) $0^\circ$ (magnetic-pole
on), (b) $45^\circ$, and (c) $90^\circ$ (along magnetic equator). The
horizontal axis shows the wavelength scaled to the terminal velocity
of the wind, $x = v/v_\infty = (\lambda-\lambda_0) c /
\lambda_0 v_\infty$. The flux is normalized to the maximum value. Note
that the asymmetry in the profile is due to occultation of a portion
of the red-shifted part of the wind by the star.}
\end{figure}

As an initial exploration of this model, we have numerically
synthesized emission line profiles of equatorial x-ray-emitting flows.
In figure \ref{stephpic} we show the results of a simple model based on a radially
directed flow confined to the magnetic equator of a hot star (using a
$\beta$ velocity law, with the flow confined to within 20
degrees of the equator).  Because stellar magnetic fields are often
tilted with respect to the rotation axis, our viewing orientation
changes with rotation phase.  In the figure we show a simulated line
profile from three different angles: 0 degrees (magnetic pole-on), 45
degrees, and 90 degrees (equator-on).  The characteristics of these
profiles are due to (the changing) projected radial velocity of the
x-ray-emitting wind, as well as (viewing-angle dependent) occultation
by the star.

Clearly, the rotation phase can affect the line profiles, which is a
conclusion we will be testing against recently obtained multi-phase
Chandra observations of $\mathrm{\mathrm{\theta}}^1$ Orionis C itself.
This star has a 45 degree angle between the magnetic and rotation axes
\cite{2002MNRAS.333...55D}, and a viewing angle that is also tilted by
45 degrees with respect to the rotation axis, giving us a full range
of viewing angles with respect to the star's magnetic axis.  Future
modeling work will include the effects of absorption by the cold wind
and also take into account more realistic and complex flow
morphologies, including spectral post processing of numerical MHD
simulations.

\end{document}